%% file: sigconf.tex
\documentclass[sigconf]{acmart}

\setcopyright{acmlicensed}
\copyrightyear{2024}
\acmYear{2024}

\acmConference[Arxiv Preprint '24]{Arxiv Preprint '24}{June 03--05,
  2024}{Arxiv}
\acmISBN{978-1-4503-XXXX-X/18/06}

\usepackage{subcaption}
\usepackage{fancyvrb}

\usepackage{pgfplots}
\usepgfplotslibrary{dateplot, statistics}
\pgfplotsset{compat=1.12}

\usepackage{xcolor}

\begin{document}

\title{Scaling CS1 Support with Compiler-Integrated Conversational AI}

%
\author{Jake Renzella}
\email{jake.renzella@unsw.edu.au}
\affiliation{%
  \institution{University of New South Wales}
  \city{Sydney}
  \state{New South Wales}
  \country{Australia}
}

\author{Alexandra Vassar}
\email{a.vassar@unsw.edu.au}
\affiliation{%
  \institution{University of New South Wales}
  \city{Sydney}
  \state{New South Wales}
  \country{Australia}
}


\author{Lorenzo Lee Solano}
\email{l.leesolano@unsw.edu.au}
\affiliation{%
  \institution{University of New South Wales}
  \city{Sydney}
  \state{New South Wales}
  \country{Australia}
}

\author{Andrew Taylor}
\email{andrewt@unsw.edu.au}
\affiliation{%
  \institution{University of New South Wales}
  \city{Sydney}
  \state{New South Wales}
  \country{Australia}
}





\renewcommand{\shortauthors}{Renzella et al.}

\begin{abstract}
This paper introduces DCC Sidekick, a web-based conversational AI tool that enhances an existing LLM-powered C/C++ compiler by generating educational programming error explanations. The tool seamlessly combines code display, compile- and run-time error messages, and stack frame read-outs alongside an AI interface, leveraging compiler error context for improved explanations.
We analyse usage data from a large Australian CS1 course, where 959 students engaged in 11,222 DCC Sidekick sessions, resulting in 17,982 error explanations over seven weeks. Notably, over 50\% of interactions occurred outside business hours, underscoring the tool's value as an always-available resource.
Our findings reveal strong adoption of AI-assisted debugging tools, demonstrating their scalability in supporting extensive CS1 courses. We provide implementation insights and recommendations for educators seeking to incorporate AI tools with appropriate pedagogical safeguards.
\end{abstract}



\received{22 July 2024}

\maketitle

\newcommand{\sidekick}{DCC Sidekick}
\newcommand{\dcc}{DCC}
\newcommand{\help}{\dcc{} Help}

\section{Introduction}

The ability to write, understand and debug code are essential components to becoming a computer programmer \cite{Cunningham2022BringingGoals, Murphy2012AbilityProgramming, Fowler2022ReevaluatingStudy}. High-quality feedback and guidance are often necessary to help with these types of skills. Explanations of code can also help students make deeper connections and build schemas of programming which improve their overall reasoning skills \cite{Marwan2019TheEnvironment, Murphy2012AbilityProgramming}. 

Traditionally, higher education models utilise, in part, teaching staff to deliver explanations to students. Drop-in help sessions, consultation hours, or online forums \cite{Doebling2021PatternsStudents} provide important opportunities for students to debug errors, clarify misconceptions, and otherwise keep students on track. With increasing enrolments and larger class sizes, the demands on help resources grows.

Over the past two years, educators have explored utilising Large Language Models (LLMs) to generate error explanations \cite{Liu2024TeachingEducation, Taylor2023DccModels, Prather2023TheEducation}, supporting time-limited educators and potentially re-imagining the way we teach and learn \cite{Grassini2023ShapingSettings}.



In this paper, we attempt to answer two main research questions:
 
 \begin{itemize}
    \item[\textbf{RQ1}] How do CS1 students engage with a compiler-integrated, conversational AI tool for programming error explanations, and to what extent do they prefer this over a non-conversational, one-shot alternative?

    \item[\textbf{RQ2}] How are generative AI explanation tools used across compile-time vs run-time programming errors?
 \end{itemize}

\section {Related Work}
There has been increased interest in generative AI over the last few years, sparked by the release of OpenAI's ChatGPT. LLMs, such as ChatGPT, are based upon transformer architecture, and at their core exploit the fact that language follows a specific orderly structure. These models are trained over large quantities of text scraped from the internet. Most LLMs are such as Codex, are also trained on millions of lines of code scraped from open-source repositories, and demonstrate capabilities in code authorship using code-writing benchmarking \cite{Chen2021EvaluatingCode}. 
 
\subsection{Large Language Models and Tools in CS1}
There are many ideas around how large language models can be applied in introductory computing to support student learning, and provide efficiencies. In introductory programming, LLMs have been used to solve simple CS1 programming exercises, with varying degrees of success \cite{Denny2023ConversingLanguage, Finnie-Ansley2022TheProgramming, Denny2024DesirableEducation, Finnie-Ansley2023MyExercises, Wermelinger2023UsingProblems}. Models have also been used to generate explanations for CS1 code. MacNeil et al.~\cite{MacNeil2022ExperiencesE-Book} generated code explanations and integrated these into an interactive e-book. They found that students viewed these explanations and perceived them to be helpful \cite{MacNeil2022ExperiencesE-Book}. Another example of using large language models in generating code explanations has been in Harvard's CS50 program, via a rubber duck persona \cite{Liu2024TeachingEducation}. This tool has been used over 50,000 times since June 2023 with an average of 15 prompts per user per day. Integrated into the programming IDE, it is able to provide explanations of code and has been positively received by students, however, no analysis of student learning was provided. Others have integrated large language models into the compiler to assist with interpreting and solving programming error messages as they occur \cite{Taylor2023DccModels}. 

Studies have evaluated the efficacy of LLM explanations, demonstrating that they produce sufficient explanations of code in introductory programming \cite{Hellas2023ExploringRequests, Kiesler2023LargeAssessments}. Tools such as the web-based CodeHelp tool, provide immediate support to students working on programming exercises and wanting assistance \cite{Liffiton2023CodeHelp:Classes}, which reduced the overall anxiety of students who were worried about asking educators for help. One limitation of CodeHelp and other error one-shot explanation tools is that they do not take advantage of the conversational capabilities of large language models, and as such do not allow for follow-up clarifications.
 
There is growing research into understanding the impact these tools have on learning outcomes. One study found that students who used Codex to learn Python could write better code than the group who did not use Codex, and demonstrated a similar understanding to the control group \cite{Kazemitabaar2023StudyingProgramming}. Prather et al.~\cite{Prather2023ItsProgrammers} found that novices struggled to understand and use these generative AI tools, and were wary about the use of such tools. Issues of over-reliance were also observed \cite{Prather2023ItsProgrammers}. Instructors report serious concerns regarding academic integrity, cheating, and a potential lack of equity, access as well as ethical objections \cite{Lau2023FromCopilot}.

\subsection{The Socratic Method}
The Socratic method has a long history of use in various educational domains to develop student critical thinking skills \cite{Knezic2010TheEducation}. This method of instruction proposes guiding student comprehension and learning through the use of guided questions, paying attention to key aspects of the target topic or concept, to help arrive at a solution. These guided questions are based on constructivist learning theories, which propose that learners build knowledge by doing rather than being told \cite{Ben-Ari2001}. This is also supported by cognitive psychology literature, which states acquiring knowledge is a function of time and conscious effort \cite{Sweller2023CognitiveLearn}. The Socratic method has been used extensively in other domains, most commonly law, but there have been few applications in computing education \cite{Tamang2021AComprehension}. One such study by Tamang et al.~\cite{Tamang2021AComprehension} used the Socratic method in teaching introductory computing skills and found that this method of guided self-explanation is more effective than free self-explanations for novice code comprehension. We seek to explore how \sidekick{} can serve as a vehicle for the Socratic method within the introductory computing course, guiding students to arrive at a solution independently, thereby improving learning outcomes.


\section {Motivation}

In our large Australian University, 3,500 students enrol in the CS1 course each year. Recent and continued growth of student enrolments is managed with a comprehensive programme of student support. In addition to scheduled class time, we run frequent scheduled help sessions for the student cohort. Over a seven week period in a previous term's help sessions, there was a total of 1,484 requests for tutor help from students, with an average time to resolution of 20 minutes. However, students were waiting on average 38 minutes before they could be assisted and only 73.9\% of queries were resolved. Waiting time increased during assessment deadline periods. Typically, unresolved student queries are directed to the online class forum, or told to wait until their scheduled class time. Such help sessions, whilst an important component of an overall experience, are still not enough to adequately meet the demand of student queries and required assistance. 

Extending previous work by Taylor et al.~which incorporated one-off LLM-generated error message explanations into educational compilers \cite{Taylor2024DccModels, Taylor2023FoundationsCompiler}, this work introduces a conversational error explanation module that allows students to engage in dialogue with an LLM about their compile- or run-time programming error.

\sidekick{} continues to utilise the full context and memory stack details provided by the compiler, while also generating a unique URL to a web interface. This interface allows students the capability to continue conversations with the LLM, providing the opportunity to guide the LLM to Socratically question the student's understanding of their program and the associated error. The tool can provide reworded, simplified, or expanded explanations depending on the student's needs.

\section{ \sidekick{} }
\sidekick{} is an extension to the existing \help{} tool, forked from the open-source repository as presented by Taylor et al.~\cite{Taylor2024DccModels}. The project utilises LLMs to produce enhanced error explanations of C/C++ programming error messages. Functioning as an add-on, our new \sidekick{} tool leverages \dcc{}'s extensive error detection and explanations system \cite{Taylor2023FoundationsCompiler} to present source code, error information and a conversational interface to an LLM.

While \help{} provides in situ (in-terminal), one-off error explanations, \sidekick{} offers conversational guidance. Students are launched into an accessible, web-based dashboard, shown in Figure~\ref{fig0}, which presents a comprehensive birds-eye-view of their source code, error message and program state, presented alongside a chat interface. Users can ask follow-up questions and receive Socratic guidance tailored to their current level of understanding.

Like \help{}, \sidekick{}'s integration with the \dcc{} compiler constructs detailed LLM prompts with no input from the student. This reduces the complexity of prompting a response, distinguishing it from contemporary tools such as CodeHelp \cite{Liffiton2023CodeHelp:Classes}, which require users to manually provide relevant information about their programming error.


\subsection{ Compiler Integration } \label{sec:integration}
To ensure a consistent learning environment, our students complete all coursework within a personal virtual environment located on the institution's privately hosted LINUX servers, which they access remotely via SSH or a VNC client. When students encounter a run-time or compile-time error using our forked version of \dcc{}, they are presented with a programming error message that contains instructions to request additional "AI-powered" clarifications by running either \verb|dcc-help| or \verb|dcc-sidekick| in the command line. The latter command displays a URL in the command line, as shown in Figure~\ref{launching}, that can be used to launch a new, private \sidekick{} session and access the web-based dashboard. 
 
The \sidekick{} command integrates with \help{}, allowing students to seamlessly transition back and forth between the two environments.

The \sidekick{} compiler integration is implemented using a small set of shell scripts that are installed globally, alongside a forked \dcc{} in the institution's environment (\autoref{fig1}). These scripts utilise existing plugin hooks within the \dcc{} compiler, caching error data and allowing \sidekick{} to be launched in response to errors or undefined behaviours. 
 
\subsection{ Graphical Interface }

\begin{figure*}[ht]
    

    \includegraphics[width=0.80\linewidth]{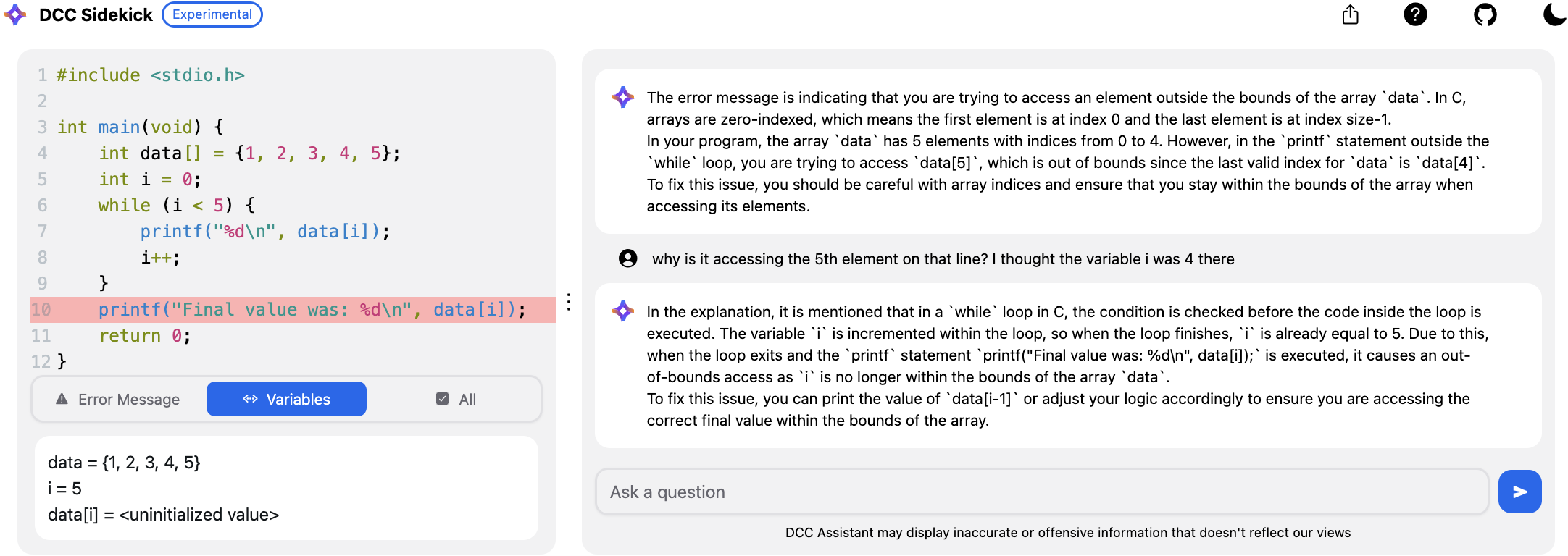}
    \caption{The \sidekick{} Interface, Where Students can Converse with an LLM about an Error they have Encountered}\label{fig0}
    \Description[ Sidekick dashboard in a web browser. ]{ The page is divided horizontally into two sections. The left-hand side displays the user's code and error. The right-hand side displays a conversation about the user's error. }
\end{figure*}

The \sidekick{} web-based dashboard is implemented in the React web framework, and prioritises accessibility for novices with a visually clear graphical interface exploring the complexities of a C error. Its separation from the locally installed integration scripts ensures student access irrespective of connection method to the institution's programming environment. Once a URL has been generated, the dashboard itself can either be accessed remotely via a VNC client, or locally on a personal device if they are instead utilising a text-only SSH connection.

When first visiting a session, \sidekick{} automatically generates and displays an initial error explanation within the chat interface, as seen in \autoref{fig0}. This aims to provide debugging guidance on the error, after which users can continue to request follow-up questions and clarifications. The dashboard also displays error message information provided directly by \dcc{}, below the user's program code. For run-time errors, this section of the dashboard dynamically displays the different categories of information across separated tabs, such as variable values, allowing users to explore the underlying complexity of the error at their own pace.

Users can also create read-only session links from within the \sidekick{} dashboard, allowing course staff to view the existing debugging information and generated explanations, streamlining the process of requesting tutor intervention.

This dashboard is backed by a privately hosted server and database, which is responsible for session creation, explanation prompting, and maintaining session data.

\subsection{ Prompting Strategy }

\begin{figure}[ht]
    \includegraphics[width=\linewidth]{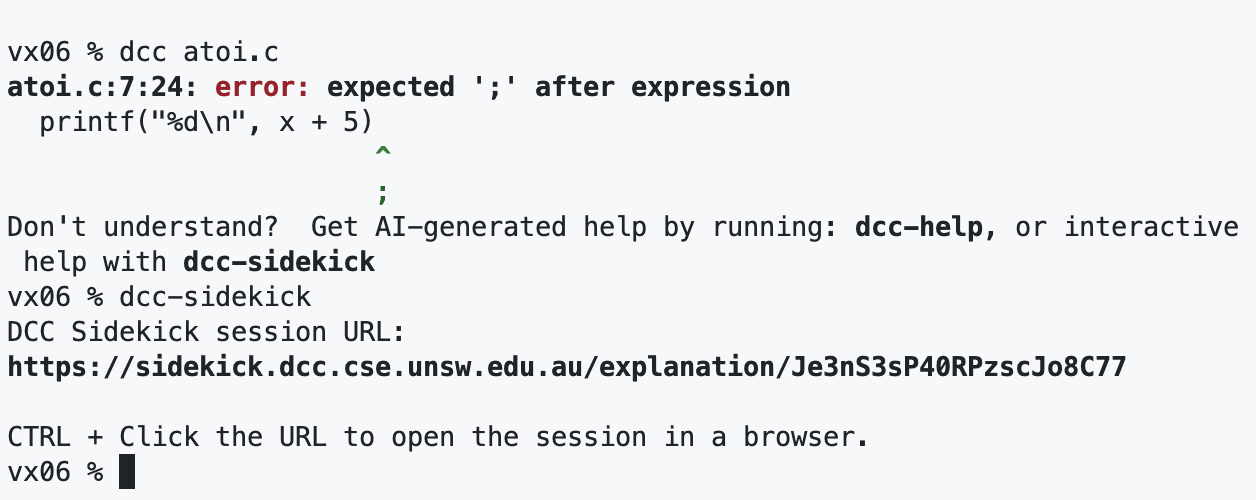}
    \caption{Launching \sidekick{} from the \dcc{} Compiler}\label{launching}
    \Description[Terminal screenshot]{Launching DCC Sidekick from the DCC Help Compiler}
\end{figure}

\begin{figure*}[ht]
    \includegraphics[width=\linewidth]{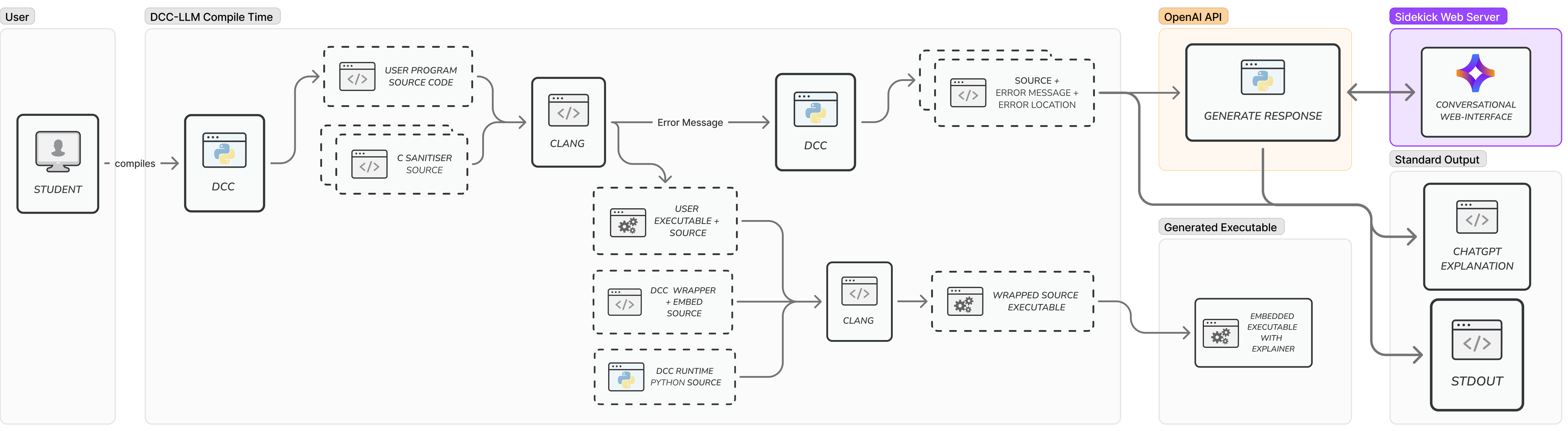}
    \caption{Diagram of Conversational \sidekick{} Generative Explanation Toolflow as a Component of the Debugging C Compiler at Compile- and Run-Time}\label{fig1}
    \Description[ Architecture diagram. ]{ The diagram displays a flowchart that shows the compilation process of DCC from the initial student's compilation request, through both GCC and Clang, generating executables and standard output which feature enhanced error explanations. These explanations and program error state are passed to OpenAI's ChatGPT 3.5 model for error explanations.}
\end{figure*}

\sidekick{} currently utilises OpenAI's ChatGPT3.5-turbo model via the chat completions API to generate both an initial error explanation and responses to follow-up queries. 
For each session, \sidekick{} maintains a conversation history to provide to the API, facilitating a coherent, relevant and accurate conversation.

The initial error explanation is generated using a context-rich prompt, derived from the program information provided by \dcc{}. The exact structure of the prompt is dependent on the type of error and the availability of data from \dcc{} \cite{Taylor2024DccModels}, but run-time errors in particular can include context such as the original error message, the user's program code, the function call stack and local variable assignments.



\begin{figure}[ht]
    \centering
    \begin{Verbatim}[frame=single, fontsize=\small]
[System]
You are a part of a programming assistant AI which is 
helping a student to debug code. Your job is to check if 
an assistant response contains programming code. 
If a response contains a code block, please rewrite the 
response so that it does not contain any code blocks, 
instead explaining the code in words.
If the original response does not contain a code block, 
then repeat the response without any changes or additional 
text.
    \end{Verbatim}
    \caption{The Guardrail System Prompt, used to Rewrite Responses that Contain Code Blocks or Solutions}
    \label{fig:rewrite-prompt}
    \Description[Fully described in the text.]{}
\end{figure}

\subsection{ Additional Guardrails }
Earlier work found that large language models often ignore instructions to not provide solutions \cite{Liffiton2023CodeHelp:Classes}, with one study observing this phenomenon in 48\% of cases \cite{Taylor2023DccModels}. To protect against this behaviour, additional guardrails based on prior work by Liffiton et al. \cite{Liffiton2023CodeHelp:Classes}, were incorporated into the \sidekick{} prompting strategy. All generated responses are filtered through an additional layer of LLM guardrails, which aim to rewrite overly prescriptive responses that contain code blocks. The system prompt for this rewriting step can be seen in Figure \ref{fig:rewrite-prompt}. 

Additionally, \sidekick{}'s standard system prompt was modified to include provisions to discourage off-topic requests and code solutions. This prompting strategy aims to encourage the generation of responses that provide valuable debugging guidance to novices, without undermining the development of their debugging skills and understanding.

Finally, the tool warns students against over-use if used frequently in a short period of time, gently reminding them that AI assistance will not be available in the final exam.

\subsection{Risks}
The tool is powered by large language models, which are innately vulnerable to hallucinations, biases, and attempts designed to break guardrails, a well-known and documented phenomenon ~\cite{Weidinger2022TaxonomyModels}. This could result in unsatisfactory or inappropriate responses, although we have not observed this behaviour directly.

\section{Methodology}
To evaluate the user adoption, impact and early markers of efficacy, we deployed \sidekick{} at our large Australian university in 2024, to a cohort of approximately 1,200 introductory programming students. An insignificant number of students in subsequent C courses may also have used the tools. 

Students were instructed on how to use \sidekick{}, alongside alternative tools like \help{}, and are prompted to create a session after any compile-time or run-time error within the university's programming environment, as described in \autoref{sec:integration}.

\subsection{Data Collection}
We collected usage data for both \sidekick{} and \help{} over the first seven weeks of an introductory programming course in C, which spanned several fundamental CS1 topics such as basic IO, control flow, arrays, and linked lists.

\subsubsection{Usage Logs}
\sidekick{} usage was measured by tracking session launches, defined as instances in which a student first creates and visits a \sidekick{} session in-browser, either in response to a programming error or to ask follow-up questions after first running \help{}. For both compile- and run-time errors, we log the source code, compiler error messages, and all generated inferences during each session, allowing us to monitor the student engagement across both \sidekick{} and \help{}.

\subsubsection{Heatmaps and Session Recordings}
Anonymised session recordings, user events and click-based heatmaps were collected during the 7-week period using the Microsoft Clarity platform\footnote{\url{https://clarity.microsoft.com}}, providing further insight into common user interaction patterns. We reflect on the usability and design of \sidekick{}'s graphical interface in \autoref{sec:lessons}.

\subsection{Data Filtering}
To comply with the relevant ethics requirements for this study, all collected student data was redacted before analysis. Identifiable features were automatically parsed and removed from usage logs and recordings. This involved removing all comments from logged student source code, in an attempt to remove occurrences of user information, such as names, student IDs and emails.

To ensure that the data accurately represents novice programmers, all uses of \sidekick{} and \help{} by university staff members, who may have initiated sessions for testing or demonstration purposes, were removed from analyses.

\section{Results}
\sidekick{} was made available to students on the first day of the term and we summarise the results of usage over the first seven weeks of the term. Overall, \sidekick{} has been used by 959 unique users within 11,222 sessions, and generating 17,982 responses. There were an average of 11.7 sessions per student.

On average, each student has spent about 4.3 minutes actively engaging with \sidekick{}. The maximum active time spent on any single interaction was 11.3 minutes. Of all these sessions, 72.8\% of sessions are from returning students, whereas 27.2\% of sessions are with new users only. This shows a high return rate and deep engagement with the tool. A total of 25.6\% of all conversations resulted in multiple inferences within one session, indicating take up of follow-up dialogue resulting from the initial error. Of all sessions launched, 22.7\% had multiple inferences within one session for compile-time errors, and 35.4\% for run-time errors. There was an increasing trend in run-time errors as the term progressed. Across all the sessions, there was an average number of 0.6 follow-up questions to each starting message. In any given session where a student asked at least one follow-up question, there was an average of 2.6 follow-up questions asked. In 12.5\% of cases, whilst a unique address was created, it was never visited. 

A total of 32,090 clicks were recorded across the sessions in the last seven weeks. The main engagement was with the conversational aspect of the tool, with 61.2\% of all clicks focused on the conversational side of the tool.

The tool was used extensively both in and out of business hours, with 44\% of use occurring within business hours (9am-5pm), and 56\% of usage occurring out of business hours (5pm-9am). Ten percent of usage occurs between the hours of midnight and 6am, when no one else is available to provide assistance to the student. 

\subsection{Usage and Adoption}

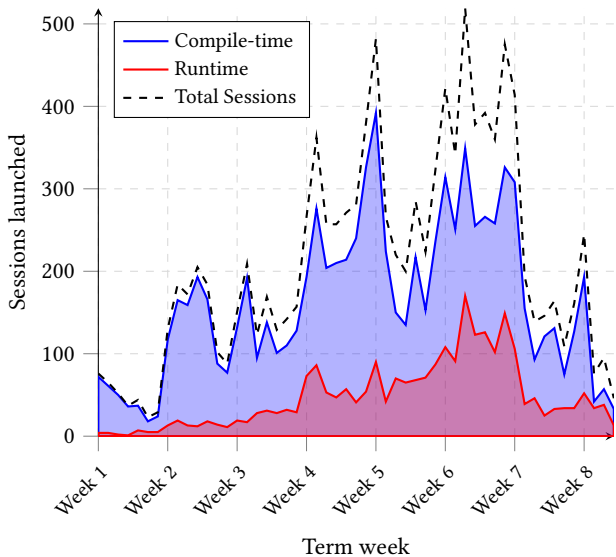
\begin{figure}[ht]
    \centering
    \input{data/help-usage/usage-chart}
    \caption{Launches Over Time of the \sidekick{} Tool, Comparing both Compile- and Run-time Errors}
    \label{fig:usage-chart}
    \Description[]{ Line graph showing Session Launched from 0 to 500 on the Y axis against Term week from Week 1 to Week 7 on the X axis. Three lines are shown for Compile-time, Runtime and Total Session. Compile-time launches are consistently higher than Runtime, with notable peaks for all session launches in Week 4 and Week 6. }
\end{figure}

\autoref{fig:usage-chart} shows \sidekick{} launches over each teaching week broken down into compile- and run-time, and total. Adoption grows steadily throughout the teaching period, with usage peaking each Monday corresponding with the weekly assessment due dates. The results show that compile-time errors contribute to the majority of the session launches, especially at the start of term. Run-time errors start to increase steadily from week 5.

While there are more compile-time error sessions of \sidekick{}, run-time sessions receive higher engagement. At compile-time, around 23\% of \sidekick{} sessions contain follow-on conversations, and approximately 35\% at run-time.

Approximately 50\% of generative AI error explanations originated from the in-terminal \help{} tool consistently throughout the term, indicating that students found value in both tools.

\section{Discussion}
Addressing RG1, adoption trends shown in \autoref{fig:usage-chart} indicate strong user acceptance. With 11,222 sessions initiated by 959 unique students, the tool demonstrates its capacity to support a large-scale introductory course. This level of engagement suggests that students find value not just in AI-enabled error explanations, but conversational debugging assistance provided by \sidekick{}.

Of the total sessions with follow-up dialogue, 1,969 were related to compile-time errors, while 899 addressed run-time issues. The predominance of follow-on dialogue in compile-time error sessions, particularly in the early weeks, aligns with the typical progression of novice programmers who often grapple with syntax and basic structural errors in the initial stages of learning. 

Addressing RG2, we observed a marked increase in run-time error assistance requests starting from week 4. This shift coincides with the introduction of more complex programming tasks in the CS1 course, particularly those involving memory management and pointers. As students tackle these more challenging concepts, they appear to be turning to \sidekick{} rather than external resources like ChatGPT, and the non-conversational, but in-terminal \help{} tool. This uptake is noteworthy, as it suggests that the benefits of the \sidekick{} conversations outweigh any perceived inconvenience of switching environments.

Usage patterns reveal a concentration of activity around assessment deadlines, peaking on Mondays when weekly assessments are typically due, and especially in week 6, which correlates with the major assignment deadline. This trend underscores the role of \sidekick{} in supporting students during high-stress periods when the teaching team may become overwhelmed by requests for assistance. Similarly, more than half the usage of the tool occurs outside of business hours, when students would not otherwise be able to get help as needed. This underscores the capability of AI-enabled tools to support teaching teams when they need it most.

The high return rate and number of follow-up questions in \sidekick{} sessions demonstrates students are meaningfully engaging to deepen their understanding and debugging techniques.

While \sidekick{} adoption is strong, it's important to note that the existing \help{} tool remains the primary source of AI-generated assistance. This inclination for in situ help suggests that immediacy and seamless integration is preferred, until a more thorough debugging session is warranted. This is a relevant finding to the wider community, as many general-purpose AI tools such as ChatGPT, CodeHelp~\cite{Liffiton2023CodeHelp:Classes}, or Harvard's CS50 forum bot~\cite{Liu2024TeachingEducation} are external to the development environment.













\subsection{Lessons Learned} \label{sec:lessons}
Our experience developing and deploying \sidekick{} presents valuable insights for tool makers, researchers, and educators:

\textbf{Integration is key}: Despite allowing external LLMs like ChatGPT, \help{} and \sidekick{} usage indicates that our approach of a) integrating with the compiler to produce more accurate responses, and b) seamlessly integrating within existing workflows, is crucial for adoption. This is important, as we can retain our pedagogical guardrails such as rate limiting and encouraging language models to provide guidance rather than solutions.

\textbf{Balancing immediacy and depth}: While \sidekick{} offers a conversational interface, the continued high usage of the simpler, terminal-based \help{} tool indicates that students value immediate, in situ assistance for certain types of errors such as simpler syntax errors in earlier weeks. Future tools should consider how to combine the benefits of both approaches.

\textbf{Investment in conversations}: When students do choose to engage with conversational debugging in \sidekick{}, they invest considerable time and effort. This is higher for run-time error cases, evidencing the role that conversational AI has in supporting more complex debugging tasks.

\section{Future Work}
We identify three avenues for future research:

\textbf{Student surveys and pedagogical impact}: We plan to interview and survey CS1 students to gain deeper insights into \sidekick{}'s efficacy and impact on long-term learning. 

\textbf{Time-series analyses}: By analysing student progress immediately following a \help{} or \sidekick{} invocation, we can quantitatively explore the impact these tools have on metrics like successful compilation and successfully passing autotests.

\textbf{Model evaluation and refinement}: We plan to assess alternative commercial and open-source LLMs, exploring on-premises hosting options and fine-tuning techniques to improve CS1 error debugging performance, and to address privacy concerns.

\section{Limitations}
We identify three limitations that impact the validity and generalisability of our findings.

Despite integrating an open-ended feedback form into the \sidekick{} tool, low response rates mean that this work lacks understanding of the qualitative aspects of \sidekick{}'s impact on the learning experience. For example, it is not clear why students choose to transition from the in-terminal \help{} responses to the conversational \sidekick{} tool. Future work such as surveys will help ascertain students' perceptions of the tool's efficacy or its influence on their problem-solving strategies. Surveys would also allow us to evidence our claims that \sidekick{} prevented or dissuaded students away from general-purpose LLMs like ChatGPT.

Secondly, our study lacks analysis of learning outcomes. While engagement metrics indicate user acceptance, they do not necessarily correlate with improved programming skills or conceptual understanding. A comparative study of academic performance between \sidekick{} users and non-users would provide more conclusive evidence of its educational value.

Finally, this study explores a single institution's CS1 course, limiting the generalisability and long-term adoption trends.

\section{Conclusion}
This paper presents \sidekick{}, a novel, compiler-integrated conversational AI tool designed to support debugging activities in a large-scale CS1 course. Our approach combines in-terminal responses via the forked \help{}, with \sidekick{}: a web-based conversational interface, providing students with flexible, context-aware debugging sessions. The high adoption rate --- 959 students initiating over 11,222 sessions in seven weeks, and significant engagement over 4.3 minutes on average, demonstrates the tool's impact. We explore behaviours across compile- and run-time errors, indicating that the nature of a programming error influences the type of AI help students engage with.

\sidekick{}'s compiler integration offers students a compelling alternative to general-purpose AI tools like ChatGPT, despite \sidekick{}'s pedagogical guardrails. By keeping students within our guided learning environment, we aim to foster genuine understanding and skill development - dissuading the use of tools that are not pedagogically aligned.

\sidekick{} demonstrates significant potential for scaling support in large programming courses, particularly in its ability to allow ongoing exploration and Socratic guidance of programming errors in a novel, compiler-integrated environment. The approach is shown to handle high volumes of student queries, which is especially valuable outside business hours and near assignment deadlines. As we continue to refine our approach to AI-generated support in introductory computer science courses, we believe that AI-assisted tools will play a crucial role in computing education. The advancement of these tools, whilst promising, needs to be executed responsibly, ensuring safe pedagogical environments that encourage learning.

\begin{acks}
We acknowledge Google for their support as part of the Google Award for Inclusion Research Program.
\end{acks}

\bibliographystyle{ACM-Reference-Format}
\bibliography{sigconf}

\end{document}

%% file: data/help-usage/usage-chart.tex
\begin{tikzpicture}
\begin{axis}[
    xlabel={Term week},
    ylabel={Sessions launched},
    date coordinates in=x,
    table/col sep=comma,
    date ZERO=2024-05-27,
    xticklabel={Week \pgfmathparse{int(\ticknum+1)}\pgfmathresult},
    xticklabel style={anchor=north east, font=\small, rotate=45},
    xtick={2024-05-27, 2024-06-03, 2024-06-10, 2024-06-17, 2024-06-24, 2024-07-01, 2024-07-08, 2024-07-15},
    x tick label as interval=false,
    legend pos=north west,
    enlarge x limits=0.05,
    ymin=0,
    axis lines=left,
    grid=both,
    grid style={dashed, gray!30},
    xlabel near ticks,
    tick align=outside,
    major tick length=0.2cm,
    minor tick length=0.1cm,
    tick pos=left,
    legend cell align=left,
    legend style={font=\small, anchor=north west}
]
\addplot+[blue, fill=blue, fill opacity=0.3, no markers, thick] table[x=date,y=compile_count] {data/help-usage/session_timeline.csv} \closedcycle;
\addlegendentry{Compile-time}
\addplot+[red, fill=red, fill opacity=0.3, no markers, thick] table[x=date,y=runtime_count] {data/help-usage/session_timeline.csv} \closedcycle;
\addlegendentry{Runtime}
\addplot+[black, thick, no markers, dashed] table[x=date,y=total_count] {data/help-usage/session_timeline.csv};
\addlegendentry{Total Sessions}

\end{axis}
\end{tikzpicture}